\documentclass[showpacs,twocolumn,aps,prl]{revtex4}
\usepackage{amsmath}
\usepackage{amsfonts}
\usepackage{amssymb}
\usepackage{amsthm}
\usepackage{graphicx}

\begin{document}

\title{Dielectric function of the semiconductor hole gas}
\author{John Schliemann}
\affiliation{Institute for Theoretical Physics, University of Regensburg,
D-93040 Regensburg, Germany}
\date{July 2010}

\begin{abstract}
The semiconductor hole gas can be viewed as the companion of the
classic interacting electron gas with a more complicated 
band structure and plays a crucial role in the understanding of
ferromagntic semiconductors. 
Here we study the dielectric function of an homogeneous hole gas in
zinc-blende III-V bulk semiconductors within random phase
approximation with the valence band being modeled by
Luttinger's Hamiltonian in the spherical approximation. In the static limit
we find a beating of Friedel oscillations between the two Fermi momenta
for heavy and light holes, while at large frequencies dramatic
corrections to the plasmon dispersion occur.
\end{abstract}
\maketitle

The interacting electron gas, combined with a homogeneous neutralizing 
background, is one of the paradigmatic systems of many-body physics
\cite{Giuliani05,Mahan00,Bruus04}.
Although obviously a grossly simplified model of a solid-state system, 
its predictions provide a good description of important properties 
of three-dimensional bulk metals and, 
in the regime of lower carrier densities, 
$n$-doped semiconductors where the electrons reside in the s-type conduction 
band.

On the other hand, in a $p$-doped zinc-blende III-V semiconductor such as GaAs,
the defect electrons or holes occupy the p-type valence band whose more
complex band structure can be expected to significantly 
modify the electronic properties. Moreover, the most intensively studied
ferromagnetic semiconductors such as Mn-doped GaAs are in fact $p$-doped
with the holes playing the key role in the occurrence of carrier-mediated 
ferromagnetism among the localized Mn magnetic moments \cite{Jungwirth06}.
Thus, such $p$-doped bulk semiconductor systems lie at the very heart of the 
still growing field of spintronics \cite{Fabian07}, and therefore it appears 
highly desirable to gain a deeper understanding of their many-body physics.

Following the above motivations, we investigate in the present letter 
the dielectric function of the homogeneous hole gas in
$p$-doped zinc-blende III-V bulk semiconductors within random phase
approximation (RPA)\cite{Giuliani05,Mahan00,Bruus04}. The single-particle
band structure of the valence band is modeled by
Luttinger's Hamiltonian in the spherical approximation \cite{Luttinger56}.
In previous work we have studied the same system using Hartree-Fock (HF)
approximation \cite{Schliemann06}. A key result here is the observation
that in a fully selfconsistent solution of the HF equations
the Coulomb repulsion among holes modifies the Fermi momenta compared to
the non-interacting situation. In particular, the selfconsistent solution of 
the HF equations is not equivalent to first-order perturbation theory as it the
case for the ordinary electron gas \cite{Giuliani05,Mahan00,Bruus04}.
Moreover, we mention recent studies of the dielectric function
in two-dimensional electron systems with spin-orbit coupling
\cite{Pletyukhov06,Badalyan09} and two-dimensional hole systems
\cite{Cheng01}. Other recent related studies have dealt with
the dielectric function of
planar graphene sheets where an effective spin is incorporated by the
sublattice degree of freedom\cite{Wunsch06,Stauber10}.

Luttinger's Hamiltonian describing heavy and light hole states around the 
$\Gamma$ in III-V zinc-blende semiconductors reads\cite{Luttinger56}
\begin{equation}
{\cal H}=\frac{1}{2m_{0}}\left(\left(\gamma_{1}+\frac{5}{2}\gamma_{2}\right)
\vec p^{2}-2\gamma_{2}\left(\vec p\cdot\vec S\right)^{2}\right)\,,
\label{Luttinger}
\end{equation}
where $m_{0}$ is the bare electron mass, $\vec p$ is the hole lattice momentum, 
and $\vec S$ are spin-$3/2$-operators.
The dimensionless Luttinger parameters $\gamma_{1}$ and $\gamma_{2}$ 
describe the valence
band of the specific material within the so-called spherical approximation
The above Hamiltonian is rotationally invariant and commutes with the 
helicity operator $\lambda=(\vec k\cdot\vec S)/k$, 
where $\vec k=\vec p/ \hbar$ 
is the hole wave vector. Thus, the eigenstates of (\ref{Luttinger})
can be chosen to be eigenstates of the helicity operator with the heavy (light)
holes corresponding to $\lambda=\pm 3/2$ ($\lambda=\pm 1/2$).
The energy dispersions are given by
$\varepsilon_{h/l}(\vec k)=\hbar^{2}k^{2}/2m_{h/l}$ where 
$m_{h/l}=m_{0}/(\gamma_{1}\mp 2\gamma_{2})$ is the effective mass of
heavy and light holes, respectively.

Combining the above single-particle Hamiltonian with Coulomb repulsion among
holes and a neutralizing background, the dielectric function within RPA
is generally given by
\begin{equation}
\varepsilon^{RPA}(\vec k,\omega)=1-V(\vec k)\chi_{0}(\vec k,\omega)\,, 
\end{equation}
where $V(\vec k)$ is the 
Fourier transform of the interaction potential, and the
free polarizability reads
\begin{eqnarray}
 \chi_{0}(\vec k,\omega) 
& = & \frac{1}{(2\pi)^{3}}\sum_{\lambda_{1},\lambda_{2}}\int d^{3}k'\Biggl[
\left|\langle\chi_{\lambda_{1}}(\vec k')|
\chi_{\lambda_{2}}(\vec k'+\vec k)\rangle\right|^{2}\nonumber\\
 & & \cdot\frac{f(\vec k',\lambda_{1})-f(\vec k'+\vec k,\lambda_{2})}
{\hbar\omega+i0-\left(\varepsilon_{\lambda_{2}}(\vec k'+\vec k)
-\varepsilon_{\lambda_{1}}(\vec k')\right)}\Biggr]\,.
\label{chi1}
\end{eqnarray}
Here $f(\vec k,\lambda)$ are Fermi functions, and the explicit form of the
four-component eigenspinors $|\chi_{\lambda}(\vec k)\rangle$ of the
Hamiltonian (\ref{Luttinger}) has been given in
Ref.~\cite{Schliemann06}. The mutual overlap of these eigenspinors
entering the above expression is a key feature of the semiconductor hole gas.

In general, an exact evaluation of the free polarizability (\ref{chi1})
is, even in the limit of zero temperature, a formidable task and
clearly more complicated than the case of the spinless electron gas.
Therefore we shall be content here with zero-temperature properties  
concentrating on the static limit, and on the regime of large frequency and
small wave vector. In the former case ($\omega=0$) an already quite tedious
calculation yields
\begin{eqnarray}
\chi_{0}(\vec k,0) & = & -\frac{m_{h}}{\pi^{2}\hbar^{2}}k_{h}
\left(1+3\left(\frac{k}{2k_{h}}\right)^{2}\right)
L\left(\frac{k}{2k_{h}}\right)\nonumber\\
 & & -\frac{m_{l}}{\pi^{2}\hbar^{2}}k_{l}
\left(1+3\left(\frac{k}{2k_{l}}\right)^{2}\right)
L\left(\frac{k}{2k_{l}}\right)\nonumber\\
 & & +\frac{3\left(\sqrt{m_{h}}+\sqrt{m_{l}}\right)^{2}}{4\pi^{2}\hbar^{2}}
\frac{k^{2}}{k_{h}+k_{l}}
L\left(\frac{k}{k_{h}+k_{l}}\right)\nonumber\\
 & & -\frac{3\left(m_{h}-m_{l}\right)^{2}}{4\pi^{2}\hbar^{2}}
\left(k_{h}-k_{l}\right)\left(1-L\left(\frac{k}{k_{h}+k_{l}}\right)\right)
\nonumber\\
 & & +\frac{3m_{h}}{2\pi^{2}\hbar^{2}}kH\left(\frac{k}{2k_{h}}\right)
+\frac{3m_{l}}{2\pi^{2}\hbar^{2}}kH\left(\frac{k}{2k_{l}}\right)\nonumber\\
 & & -\frac{3\left(m_{h}+m_{l}\right)^{2}}{4\pi^{2}\hbar^{2}}
kH\left(\frac{k}{k_{h}+k_{l}}\right)\,,\label{chi2}
\end{eqnarray}
where $k_{h/l}=\sqrt{2m_{h/l}\varepsilon_{F}/ \hbar^{2}}$ are the Fermi wave numbers
for heavy and light holes at Fermi energy $\varepsilon_{F}$. The so-called
Lindhard correction $L$ is given by
\begin{equation}
L(x)
=\left(\frac{1}{2}+\frac{1-x^{2}}{4x}\ln\left|\frac{1+x}{1-x}\right|\right)\,,
\end{equation}
and the function $H$ is defined as
\begin{eqnarray}
H(x) & = & \frac{1}{2}\int_{0}^{1/x}dy\frac{1}{y}\ln\left|\frac{1+y}{1-y}\right|
\nonumber\\
 & = & \left\{
\begin{array}{ll}
\frac{\pi^{2}}{4}-\sum_{n=0}^{\infty}\frac{x^{2n+1}}{(2n+1)^{2}} & |x|\leq 1 \\
\sum_{n=0}^{\infty}\frac{\left(\frac{1}{x}\right)^{2n+1}}{(2n+1)^{2}} & |x|\geq 1
\end{array}
\right.\,.
\end{eqnarray}
Remarkably, one can express the polarizability entirely in terms of the 
arguments $k/2k_{h}$, $k/2k_{l}$, and $k/k_{h}+k_{l}$ with the latter one being
the harmonic mean of the two former. In the limit $m_{h}=m_{l}$
(i.e. $k_{h}=k_{l}=:k_{F}$) one obtains the usual result
$\chi_{0}(\vec k,0)=-D(\varepsilon_{F})L(k/2k_{F})$ for charge
carriers without spin-orbit coupling where $D(\varepsilon)$
is the density of states \cite{Simion05}. The full polarization (\ref{chi2}
at $m_{h}\neq m_{l}$, however, has a clearly mor complicted structure.

On the other hand, considering
Coulomb repulsion, $V(\vec k)=e^{2}/\varepsilon_{r}\varepsilon_{0}k^{2}$,
and using the long-wave approximation 
$\chi_{0}(\vec k,0)\approx\chi_{0}(0,0)$
leads to the usual Thomas-Fermi (TF) screening,
$\varepsilon^{RPA}(\vec k,0)\approx 1-k^{2}_{TF}/k^{2}$ with
$k^{2}_{TF}=(e^{2}/\varepsilon_{r}\varepsilon_{0})3n/2\varepsilon_{F}$.
Here $\varepsilon_{r}$ is the background dielectric constant taking into
account screening by deeper bands, and the hole density is given by
$n=n_{h}+n_{l}$, $n_{h/l}=k_{h/l}^{3}/3\pi^{2}$. 

\begin{figure}
\begin{center}
\includegraphics[width=8.5cm]{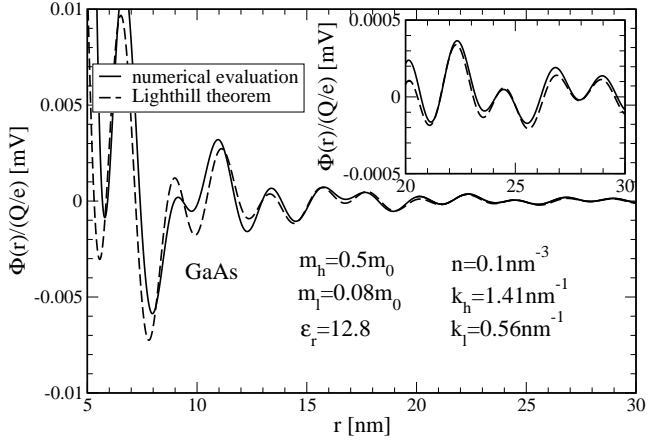}
\end{center}
\caption{Friedel oscillations resulting from a  numerical
evaluation of the Fourier integral (\ref{fourier}), and estimated
via the Lighthill theorem (cf.  Eq.~(\ref{friedel}))
for $p$-doped GaAs with a hole density of $n=10^{20}{\rm cm}^{-3}$.
The inset shows the data at larger distances on a smaller scale.
\label{fig1}}
\end{figure}

The full screened potential of
a pointlike probe charge $Q$ is given by
\begin{equation}
\Phi(\vec r)=\frac{1}{(2\pi)^{3}}\int d^{3}k
\frac{\frac{Q}{\varepsilon_{r}\varepsilon_{0}k^{2}}}
{\varepsilon^{RPA}(\vec k)}e^{i\vec k\vec r}
\label{fourier}
\end{equation}
whose asymptotic behavior is determined by the singularities of the integrand
and its derivatives \cite{Lighthill58}. Here the first derivative has 
singularities at $k=2k_{h}$ and $k=2k_{l}$ while at
$k=k_{h}+k_{l}$ all singular contributions cancel out. As a result, the
Lighthill theorem \cite{Lighthill58} yields for large distances $r$
\begin{equation}
\Phi(r)\approx\frac{m_{h}}{m_{0}}\phi_{\infty}(2k_{h},r)
+\frac{m_{l}}{m_{0}}\phi_{\infty}(2k_{l},r)
\label{friedel}
\end{equation}
where
\begin{equation}
\phi_{\infty}(q,r)=\frac{Q}{4\pi\varepsilon_{0}a_{0}}\frac{2}{\pi}
\frac{1}{\left(\varepsilon_{r}\varepsilon^{RPA}(q)\right)^{2}}
\frac{\cos(qr)}{(qr)^{3}}
\end{equation}
and $a_{0}=4\pi\varepsilon_{0}\hbar^{2}/m_{0}e^{2}$ being the usual Bohr radius.
Thus, we observe a beating of Friedel oscillations between the two 
wave numbers $2k_{h/l}$. Note that, differently form the expression
for the dielectric function itself, the wave number $k=k_{h}+k_{l}$ does not
occur in the Friedel oscillations since the non-interacting ground state
of the hole gas has singularities in the occupation numbers at
$k=k_{h/l}$ but not at $k=(k_{h}+k_{l})/2$. Fig.~\ref{fig1} shows the Friedel 
oscillations according to Eq.~(\ref{friedel}) along with a numerical
evaluation of the full Fourier integral (\ref{fourier})
for $p$-doped GaAs with a hole density of $n=10^{20}{\rm cm}^{-3}$, which is
a very typical value for Mn-doped GaAs \cite{Jungwirth06}.
One might argue whether one should replace the Fermi momenta
$k_{h/l}$ with renormalized values arising from a fully self-consistent
solution to the HF equations. However, at large densities
this renormalization becomes negligible \cite{Schliemann06}.

The beating of Friedel oscillations illustrated in the figure
is a peculiarity of the holes residing in the p-type valence band
and should be observable via similar scanning tunneling microscopy techniques
as used in metals \cite{Crommie93} and $n$-doped semiconductors
\cite{Kanisawa01}. Moreover, as theoretical studies have revealed, such 
oscillations can have a profound impact on the magnetic properties of
ferromagnetic semiconductors \cite{Schliemann02,Fiete05}.
Moreover, Fig.~\ref{fig1} shows the amazing accuracy of the
asymptotic expression (\ref{friedel}) obtained from the Lighthill theorem.

Let us now turn to the regime of large frequencies and small wave vectors.
Following Ref.~\cite{Mahan00} we expand the denominators in Eq.~(\ref{chi1})
assuming $\hbar\omega>>\varepsilon_{h/l}(\vec k)$ and 
$\hbar\omega>>(\hbar k_{h/l}/m_{h/l})\hbar k$. Within the two leading orders 
one finds
\begin{eqnarray}
 &  & \varepsilon^{RPA}(\vec k,\omega)=1-\frac{1}{\omega^{2}}
\frac{e^{2}}{\varepsilon_{r}\varepsilon_{0}}\frac{1}{6\pi^{2}}
\left(\frac{1}{m_{h}}+\frac{1}{m_{l}}\right)
\left(k_{h}^{3}+k_{l}^{3}\right)\nonumber\\
 & & \qquad-\frac{1}{\omega^{4}}\frac{e^{2}\hbar^{2}}
{\varepsilon_{r}\varepsilon_{0}\pi^{2}}
\frac{1}{2}\left(\frac{1}{m_{h}^{3}}+\frac{1}{m_{l}^{3}}\right)
\Biggl[\frac{1}{5}k^{2}\left(k_{h}^{5}+k_{l}^{5}\right)\nonumber\\
 & & \qquad\qquad+\frac{1}{12}k^{4}\left(k_{h}^{3}+k_{l}^{3}\right)\Biggr]
\nonumber\\
& & \qquad-\frac{1}{\omega^{4}}\frac{e^{2}\hbar^{2}}
{\varepsilon_{r}\varepsilon_{0}\pi^{2}}
\Biggl[-\frac{1}{56}\left(\frac{1}{m_{h}^{3}}-\frac{1}{m_{l}^{3}}\right)
\left(k_{h}^{7}-k_{l}^{7}\right)\nonumber\\
 & & \qquad\qquad+\frac{3}{56}\left(\frac{1}{m_{h}}-\frac{1}{m_{l}}\right)^{2}
\left(\frac{k_{h}^{7}}{m_{h}}+\frac{k_{l}^{7}}{m_{l}}\right)\nonumber\\
 & & \qquad\qquad+\frac{3}{28}\left(\frac{1}{m_{h}}-\frac{1}{m_{l}}\right)
\left(\frac{k_{h}^{7}}{m_{h}^{2}}-\frac{k_{l}^{7}}{m_{l}^{2}}\right)\nonumber\\
 & & \qquad\qquad+\frac{21}{200}k^{2}
\left(\frac{1}{m_{h}^{3}}-\frac{1}{m_{l}^{3}}\right)
\left(k_{h}^{5}-k_{l}^{5}\right)\nonumber\\
 & & \qquad\qquad-\frac{3}{40}k^{2}\left(\frac{1}{m_{h}}-\frac{1}{m_{l}}\right)
\left(\frac{k_{h}^{5}}{m_{h}^{2}}-\frac{k_{l}^{5}}{m_{l}^{2}}\right)\Biggr]
\label{high}
\end{eqnarray}
For $m_{h}=m_{l}$ the first three lines of the above expression reproduce again
the standard textbook result \cite{Mahan00} while all other terms vanish in 
this limit. On the other hand, if $m_{h}\neq m_{l}$, one has contributions
in order $1/\omega^{4}$ that are {\em independent of the wave vector} $\vec k$.
Such terms are absent in the case of the standard electron gas where
the contributions of order $1/\omega^{2n}$ are at least of order
$k^{2n-2}$ in the wave vector \cite{Mahan00}. The technical reason
why such contributions are present for the hole gas is that the expression
$\varepsilon_{\lambda_{2}}(\vec k'+\vec k)-\varepsilon_{\lambda_{1}}(\vec k')$
in Eq.(\ref{chi1}) contains for $|\lambda_{1}|\neq|\lambda_{2}|$ an additive 
term which is independent of $k$ (and vanishes for $m_{h}=m_{l}$).
These {\em prima vista} unexpected contributions to the high-frequency 
expansion of the dielectric function will also occur in even higher orders.
However, even in the two leading orders given in Eq.~(\ref{high}),  
they strongly modify the plasmon dispersion determined by 
$\varepsilon^{RPA}(\vec k,\omega(k))=0$ which can be expressed as
\begin{eqnarray}
  & & \omega^{2}(k)=\left(\omega_{p}^{(0)}\right)^{2}
\Biggl[\frac{1}{2}+\frac{1}{2}\Bigl[1+4\Bigl(u\left(n^{1/3}a_{0}\right)
\nonumber\\
 & & \qquad\qquad+\left(v+w\right)\frac{(ka_{0})^{2}}{n^{1/3}a_{0}}\Bigr)\Bigr]^{1/2}
\Biggr]+{\cal O}\left(k^{4}\right)
\label{plasma1}\\
  & & \approx\left(\omega_{p}^{(0)}\right)^{2}\left(1+u\left(n^{1/3}a_{0}\right)
+\left(v+w\right)\frac{(ka_{0})^{2}}{n^{1/3}a_{0}}\right)\label{plasma2}
\end{eqnarray}
where the zero-order plasma frequency is given by\cite{note}
\begin{equation}
\left(\omega_{p}^{(0)}\right)^{2}
=\frac{e^{2}}{\varepsilon_{r}\varepsilon_{0}}\frac{n}{2}
\left(\frac{1}{m_{h}}+\frac{1}{m_{l}}\right)\,,
\label{plasma3}
\end{equation}
and the dimensionless and density-independent coefficients 
$u$, $v$, $w$ are given by 
\begin{eqnarray}
 & & u=\frac{Q\left(m_{h},m_{l}\right)}
{\left(3\pi^{2}\right)^{1/3}\left(m_{h}^{3/2}+m_{l}^{3/2}\right)^{2/3}}\nonumber\\
 & & \quad\times\Biggl[
-\frac{3}{14}\left(\frac{1}{m_{h}^{3}}-\frac{1}{m_{l}^{3}}\right)
\left(m_{h}^{7/2}-m_{l}^{7/2}\right)\nonumber\\
 & & \qquad\qquad+\frac{9}{14}\left(\frac{1}{m_{h}}-\frac{1}{m_{l}}\right)^{2}
\left(m_{h}^{5/2}+m_{l}^{5/2}\right)\nonumber\\
 & & \qquad\qquad+\frac{9}{7}\left(\frac{1}{m_{h}}-\frac{1}{m_{l}}\right)
\left((m_{h}^{3/2}-m_{l}^{3/2}\right)\Biggr]\,,
\label{u}
\end{eqnarray}
\begin{equation}
v=Q\left(m_{h},m_{l}\right)\frac{2}{5\pi^{2}}
\left(\frac{1}{m_{h}^{3}}+\frac{1}{m_{l}^{3}}\right)
\left(m_{h}^{5/2}+m_{l}^{5/2}\right)\,,
\end{equation}
\begin{eqnarray}
 & & w=Q\left(m_{h},m_{l}\right)\Biggl[\frac{21}{50\pi^{2}}
\left(\frac{1}{m_{h}^{3}}-\frac{1}{m_{l}^{3}}\right)
\left(m_{h}^{5/2}-m_{l}^{5/2}\right)\nonumber\\
 & & \qquad\qquad-\frac{3}{10\pi^{2}}
\left(\frac{1}{m_{h}}-\frac{1}{m_{l}}\right)
\left(m_{h}^{1/2}-m_{l}^{1/2}\right)\Biggr]
\end{eqnarray}
with the common prefactor
\begin{equation}
Q\left(m_{h},m_{l}\right)\frac{\frac{\varepsilon_{r}}{4\pi}m_{0}}
{\left(\frac{1}{m_{h}}+\frac{1}{m_{l}}\right)^{2}}
\frac{(3\pi^{2})^{5/3}}{\left(m_{h}^{3/2}+m_{l}^{3/2}\right)^{5/3}}\,.
\end{equation}
Clearly, the coefficients $u$ and $w$ vanish for $m_{h}=m_{l}$ while
from $v$ one recovers usual textbook result for an electron gas without
spin-orbit coupling \cite{Mahan00}. By expanding the square root in 
Eq.~(\ref{plasma1}) we have neglected higher contributions both in
wave vector and in the density parameter
$n^{1/3}a_{0}\propto(\varepsilon_{F}/\hbar\omega_{p}^{(0)})^{2}$ which
is consistent with considering only the first two leading orders in 
Eq.~(\ref{high}). In fact, for usual $p$-doped bulk semiconductors
$n^{1/3}a_{0}$ is small, and to consistently obtain contributions
to the plasmon dispersion being of higher order in the density
would require to extend the expansion (\ref{high}) also to higher orders,
which is computationally increasingly tedious and will lead to even
lengthier expressions.
\begin{table}
\begin{tabular}{c|c|c|c|c|c|c|c|}
  & $\frac{m_{h}}{m_{0}}$ &  $\frac{m_{l}}{m_{0}}$ & $\varepsilon_{r}$ 
& $\frac{m_{l}}{m_{h}}$ & $u$ & $v$ & $w$\\ 
\hline
AlAs & 0.47 & 0.18 & 10.0 & 0.38 & 17.7 & 21.5 & -16.3 \\
AlSb & 0.36 & 0.13 & 12.0 & 0.36 & 49.7 & 37.1 & -29.5 \\
GaAs & 0.5 & 0.08 & 12.8  & 0.16 & 195.4 & 99.4 & -100.5 \\
InAs & 0.5 & 0.026 & 14.5 & 0.052 & 861.4 & 451.9 & -473.1 \\
InSb & 0.2 & 0.015 & 18.0 & 0.075 & 1796.9 & 919.2 & -958.8 \\
\end{tabular}
\caption{Material parameter and
coefficients $u$, $v$, $w$ of the plasmon dispersion 
(\ref{plasma2}) for various III-V semiconductors.
\label{table1}}
\end{table}
Note that the dispersion coefficients $u$, $v$, $w$ depend entirely on 
material parameters. In table \ref{table1} we have listed their numerical values
for several prominent III-V semiconductor systems. As seen there, the
coefficient $u$ is remarkably large leading to a substantial enhancement
of the long-wavelength plasma frequency 
$\omega^{2}(0)=(\omega_{p}^{(0)})^{2}(1+u(n^{1/3}a_{0})$, 
even at small densities, compared to the naive guess
$\omega^{2}(0)\approx(\omega_{p}^{(0)})^{2}$.
On the other hand, $v$ and  $w$  differ in sign
and are of quite similar magnitude resulting in a dramatic
flattening of the plasma dispersion compared to the standard case
$m_{h}=m_{l}$ where $w$ vanishes. Moreover, the sum $v+w$ can even become negative
leading to a plasmon dispersion bending downwards around zero wave vector. 
In fact the sign of $v+w$ is entirely determined by the ratio
$m_{l}/m_{h}$ where negative values occur for $m_{l}/m_{h}\lesssim 0.18$.
Remarkably, GaAs lies very close this threshold showing already
such a qualitative change in the plasmon dispersion. This trend is further 
enhanced in the cases of InAs and InSb.

In summary, we have studied 
the dielectric function of the homogeneous hole gas in
$p$-doped zinc-blende III-V semiconductors. In the static limit we predict
additional beatings of the Friedel oscillations which should be
experimentally detectable via state-of-the-art scanning tunneling microscopy.
At high frequencies and
small wave vectors the plasmon dispersion gets dramatically altered compared to
the textbook case of the usual electron gas.

I thank J. Repp for useful discussions and 
Deutsche Forschungsgemeinschaft for support via SFB 689.

\end{document}